\documentclass[11pt,twoside]{article}
\usepackage{asp2010}

\resetcounters

\begin{document}

\title{The Effelsberg--Bonn \ion{H}{i} Survey: Milky Way data}
\author{B. Winkel, J. Kerp, P.\,M.\,W. Kalberla, and N. Ben Bekhti}
\affil{Argelander-Institute for Astronomy (AIfA), University of Bonn,\\  Auf dem H\"{u}gel 71, D-53121 Bonn, Germany}

\begin{abstract}
Since autumn 2008 an L-band 7-Feed-Array is operated for astronomical science at the 100-m radio telescope at Effelsberg. This receiver is used to perform an unbiased, fully sampled \ion{H}{i} survey of the whole northern hemisphere observing both the galactic and extragalactic sky in parallel --- the Effelsberg-Bonn \ion{H}{i} survey (EBHIS). We present first results based on the Milky Way data. Up to now two larger coherent regions were mapped each covering about $2000\,\mathrm{deg}^2$. One of these fields covers the northern part of the high-velocity cloud complex GCN. With the better angular resolution of the EBHIS we resolve the previously detected clouds into isolated compact clumps and find a linewidth--$v_\mathrm{lsr}$ relation giving hints on an interaction of accreting material with the Milky Way halo.
\end{abstract}

\section{Introduction}
Performing blind \ion{H}{i} surveys opens the door to many interesting science cases. A prime example is the measurement of the \ion{H}{i} mass function (HIMF) which is also of cosmological importance. Today the largest database to address the scientific questions related to the HIMF is based on the  \ion{H}{i} Parkes All-Sky Survey \citep[HIPASS;][]{barnes2010} containing more than 4000 galaxies. Unfortunately, the sensitivity of HIPASS was not good enough to study the low-mass end or environmental/evolutionary effects on the HIMF. Hence, several projects are underway to enlarge the number of extragalactic \ion{H}{i} sources. 

Using the ALFA receiver at the Arecibo telescope the Arecibo Legacy Fast ALFA Survey \citep[ALFALFA;][]{giovanelli2005} is currently carried out, a blind survey of the sky accessible to Arecibo (7000 deg$^2$). Another large survey, the Effelsberg-Bonn \ion{H}{i} survey (EBHIS), is performed by our group utilizing a new L-band 7-Feed-Array at the 100-m radio telescope at Effelsberg. With this instrument we will map the complete northern hemisphere out to a distance of about 230\,Mpc. Both surveys, ALFALFA and EBHIS are designed to be an order of magnitude deeper than HIPASS. For EBHIS using state-of-the-art FPGA spectrometers provides high dynamic range and makes short dump times of the \ion{H}{i} spectra possible. This allows to substantially reduce the impact of radio frequency interference (RFI). The Milky Way data are corrected for the stray-radiation bias which warrants a main-beam efficiency of 99\%. 

Due to the large number of spectral channels of the FPGA spectrometers the resulting frequency resolution allows to simultaneously use the galactic velocity data to provide a northern-sky complement to the recently performed Galactic All-Sky Survey \citep[GASS;][]{mcclure2009,kalberla2010} on the southern hemisphere. This will lead to an ultimate successor of the Leiden/Argentine/Bonn survey \citep[LAB;][]{kalberla05} having much higher angular resolution ($9\arcmin$ compared to $30\arcmin$) on a fully sampled grid while LAB used beam-by-beam sampling. EBHIS as well as GASS allow to study the Milky Way disk and halo with great detail. While the thin disk was observed in the framework of the various galactic plane surveys \citep[CGPS, SGPS, and VGPS;][]{taylor2003,mcclure2005,stil2006} compact structures at higher latitudes were often simply overlooked in earlier low-resolution surveys as the LAB. A prime example is the discovery of so-called ultra-compact high-velocity clouds \citep[UCHVC;][]{bruens2004,hoffman2004}. One goal of EBHIS is to find out whether UCHVCs are frequent and, if so, study their statistical properties. Another key project is the complex interplay of the disk and halo. Processes like galactic fountains and winds are thought to inject material into the halo, while at the same time gas (e.g., from tidal interactions) is accreted onto the disk. The data can be complemented with metal absorption line studies against quasars \citep[][see also this book]{benbekhti2008} providing low-column density measurements on random sight lines.

\section{Data}\label{secdata}
The receiving system used for EBHIS provides 16384 spectral channels over a bandwidth of 100\,MHz resulting in a channel separation of about $1.3\,\mathrm{km\,s}^{-1}$. Spectral dumps are recorded every 500\,ms which yields a relatively high data rate of 5\,GB\,hr$^{-1}$ putting tight constraints on the data reduction software in terms of computing efficiency and performance. 

Extensive test measurements showed the stability of the system. Typical system temperatures lie in the range between 22 and 39\,K depending strongly on elevation angle and to some extent on local weather conditions and season. Offset feeds generally have slightly higher values than the central beam. A detailed description of the receiving system and data reduction can be found in \citet{winkel2010}. 

\section{Survey strategy and current sky coverage}
EBHIS is carried out in three major steps. First, several test measurements were scheduled, mapping smaller portions of the sky to check the instrument and software. Currently, the complete northern hemisphere is observed down to $\mathrm{decl.}> -5\deg$ yielding an RMS noise level of $\lesssim90\,\mathrm{mK}$. Finally, the Sloan Digital Sky Survey \citep[SDSS;][]{adelman08} area will be mapped via multiple coverages for a total integration time of 10\,minutes per position. We subdivide the sky into
$5\times5$ $\mathrm{deg}^2$ fields which are measured with on-the-fly R.A.--Decl. scanning using a tangential plane projection. This will lead to a homogeneous noise distribution all over the sky. The hexagonal feed pattern is rotated by $19\deg$ with respect to the scanning direction such that the sampled scan lines are equidistantly spaced. To keep this feed angle fixed in the R.A.--Decl. system, the dewar must be rotated according to the parallactic angle.
\begin{figure}[!t]
\centering
\includegraphics[width=0.8\textwidth,bb=68 77 406 264,clip=]{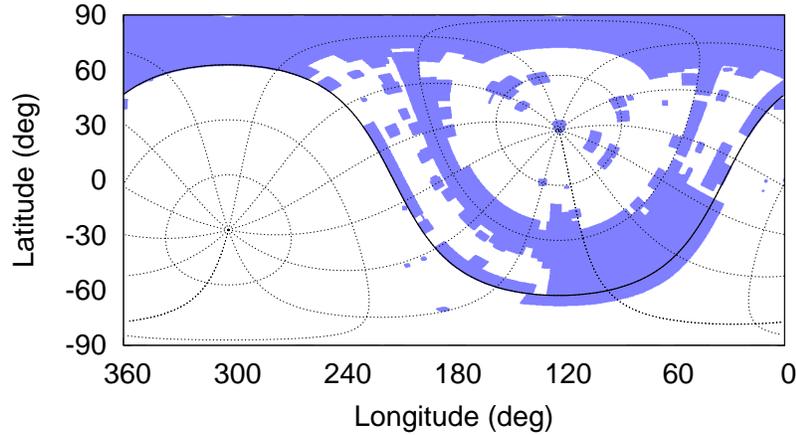}
\caption{Current sky coverage of the EBHIS.}
\label{figskycoverage}
\end{figure}

Fig.\,\ref{figskycoverage} shows the current sky coverage as of June 2010. The first coverage of the northern hemisphere is expected to be completed by Spring 2011. Data will be made publicly available in the near future on \url{http://www.astro.uni-bonn.de/hisurvey/}.

\section{First results}\label{secresults}
As displayed in Fig.\,\ref{figskycoverage} we already mapped two larger coherent areas on the sky each having about 2\,000\,deg$^2$. The first is toward the tip of the Magellanic Stream and the second covers the north galactic pole. The former region partly covers the HVC complex Galactic Center Negative (GCN) as defined by \citet{wakker1991}. 

\subsection{A revised picture of complex GCN}
\begin{figure}[!t]
\centering
\includegraphics[scale=0.4,bb=17 131 772 548,clip=]{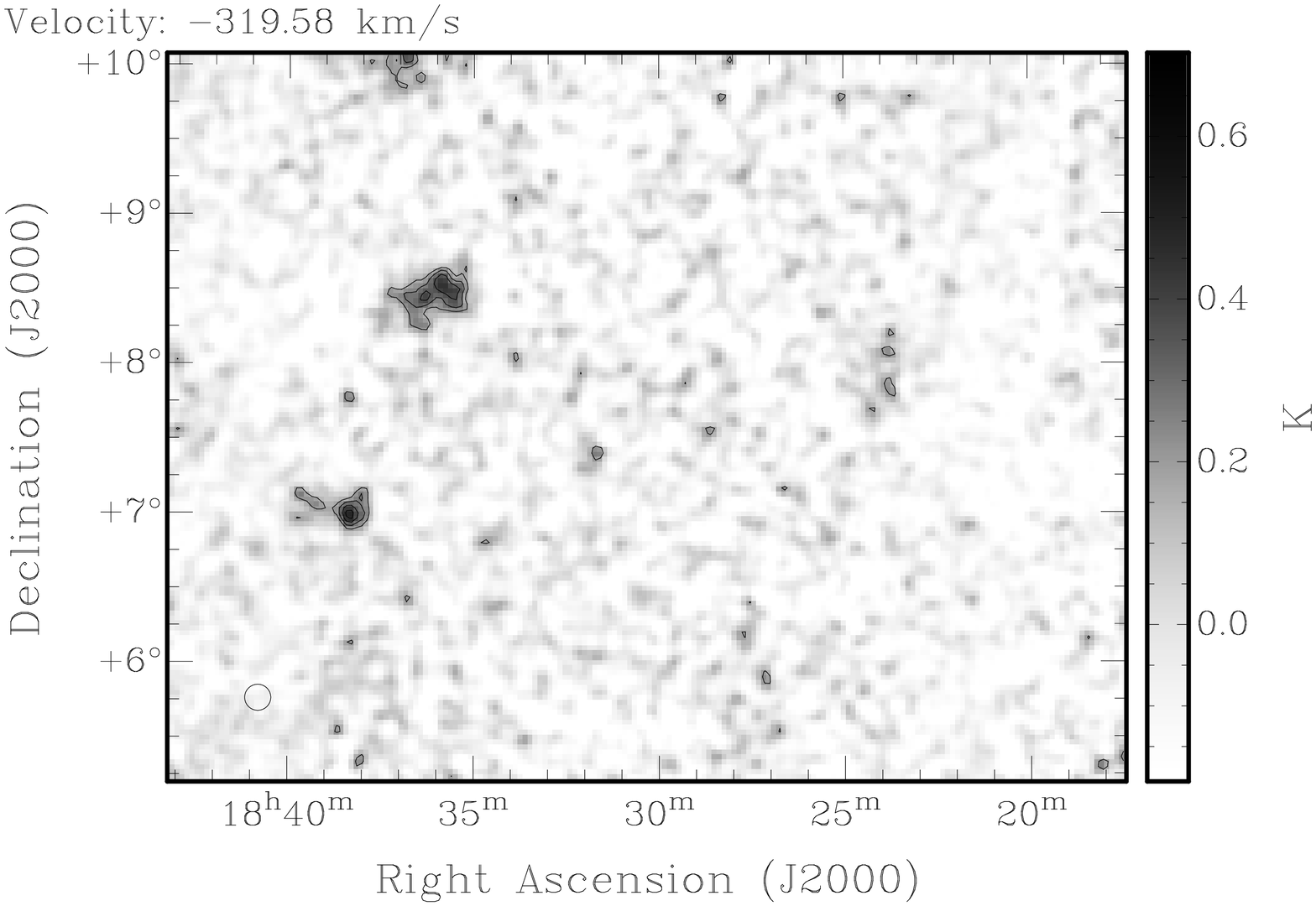}\hbox{\raise12.8em\vbox{\moveleft22.4em\hbox{$v_\mathrm{lsr}=-320\,\mathrm{km\,s}^{-1}$}}}
\includegraphics[scale=0.4,bb=17 58 772 548,clip=]{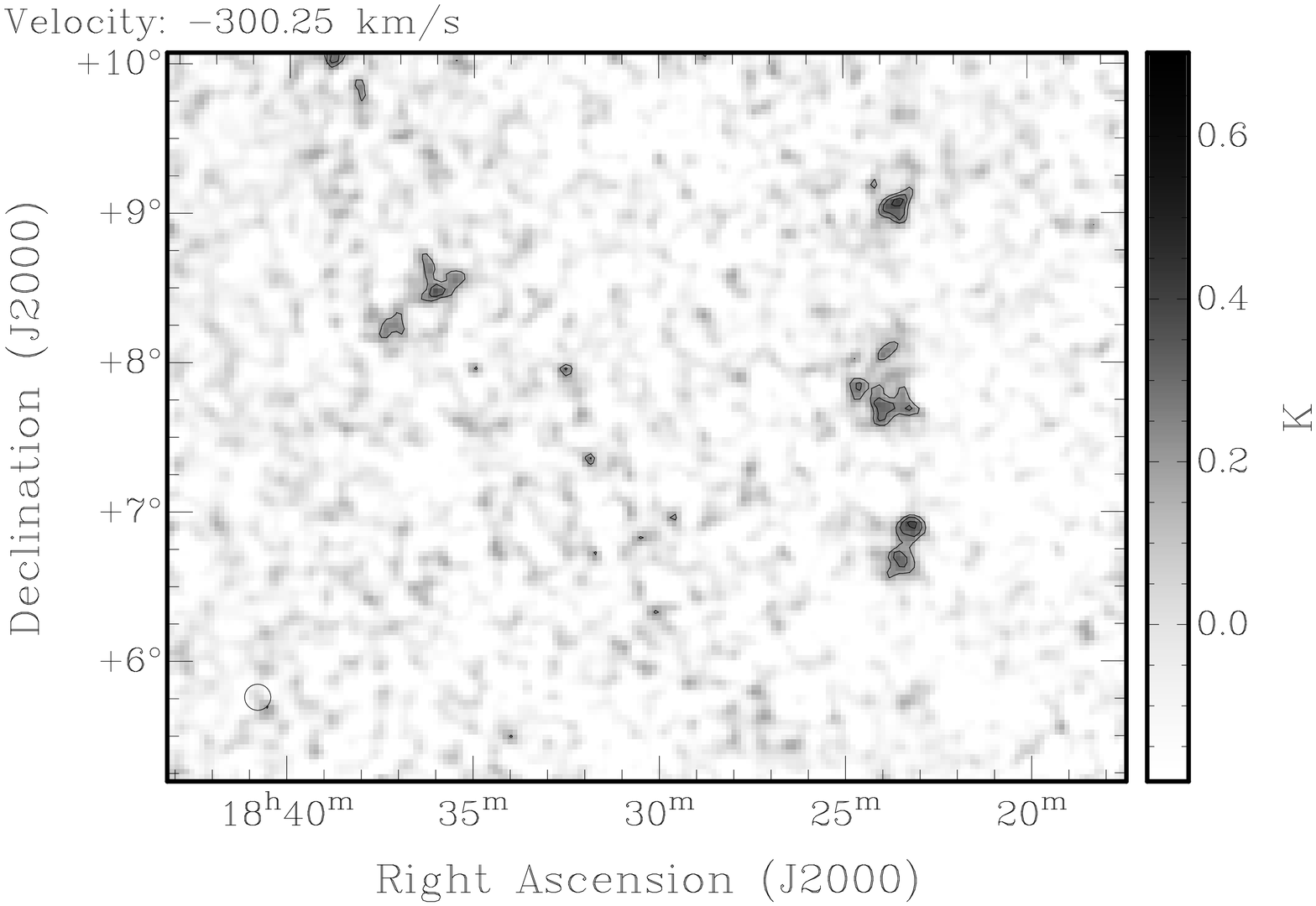}\hbox{\raise15.6em\vbox{\moveleft22.4em\hbox{$v_\mathrm{lsr}=-300\,\mathrm{km\,s}^{-1}$}}}
\caption{Two velocity planes showing some of the detected clumps in HVC complex GCN, previously unresolved. Contours start at $2\sigma_\mathrm{rms}$ in steps of $1\sigma_\mathrm{rms}=90\,\mathrm{mK}$. The size of the beam ($9\arcmin$) is indicated by the circle in the lower left.}
\label{fighvcassociationplanes}
\end{figure}
\begin{figure}[!t]
\centering
\includegraphics[width=0.98\textwidth]{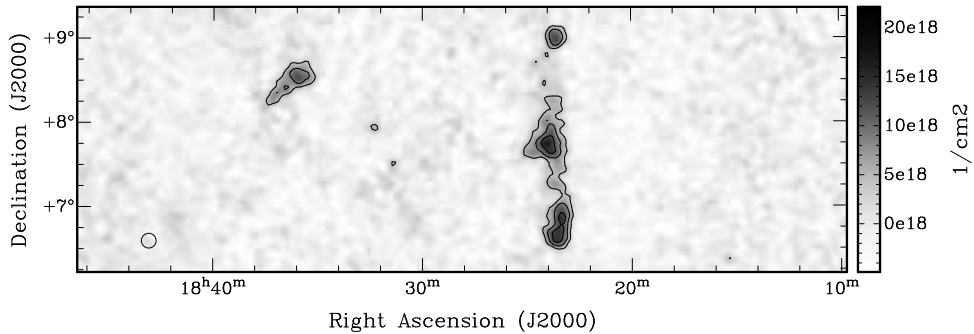}
\caption{Column density map (integrated over the velocity interval $-310\ldots-276\,\mathrm{km\,s}^{-1}$) of some of the clumps showing that down to the sensitivity level of the EBHIS no extended diffuse emission is detected in the vicinity of the compact clumps. Contours start at $3\sigma_\mathrm{rms}$ in steps of $3\sigma_\mathrm{rms}=4.8\cdot10^{18}\,\mathrm{cm}^{-2}$. Some clumps show head-tail-like morphologies or form filamentary structures, although in the latter case the individual clumps are still quite well separated in velocity by more then $10\,\mathrm{km\,s}^{-1}$.}
\label{fighvcnhi}
\end{figure}

\begin{figure}[!t]
\centering
\includegraphics[width=0.45\textwidth,clip=,bb=68 58 400 291]{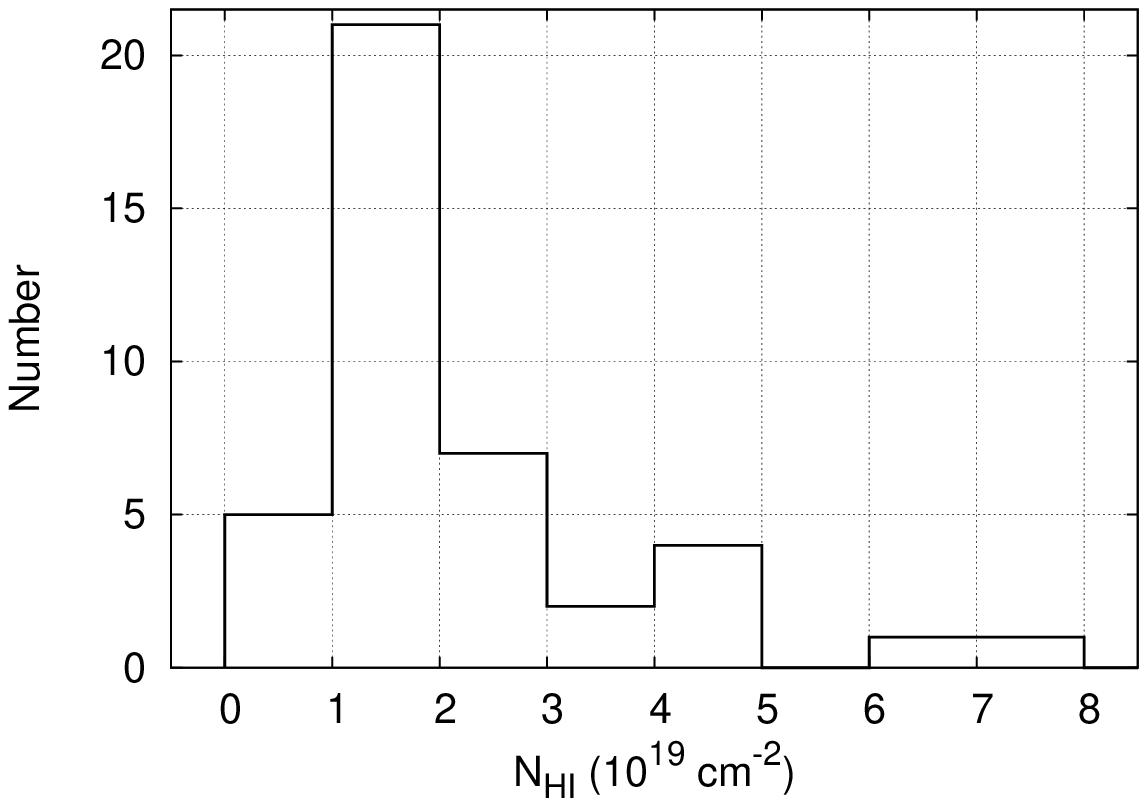}\qquad
\includegraphics[width=0.45\textwidth,clip=,bb=61 58 400 291]{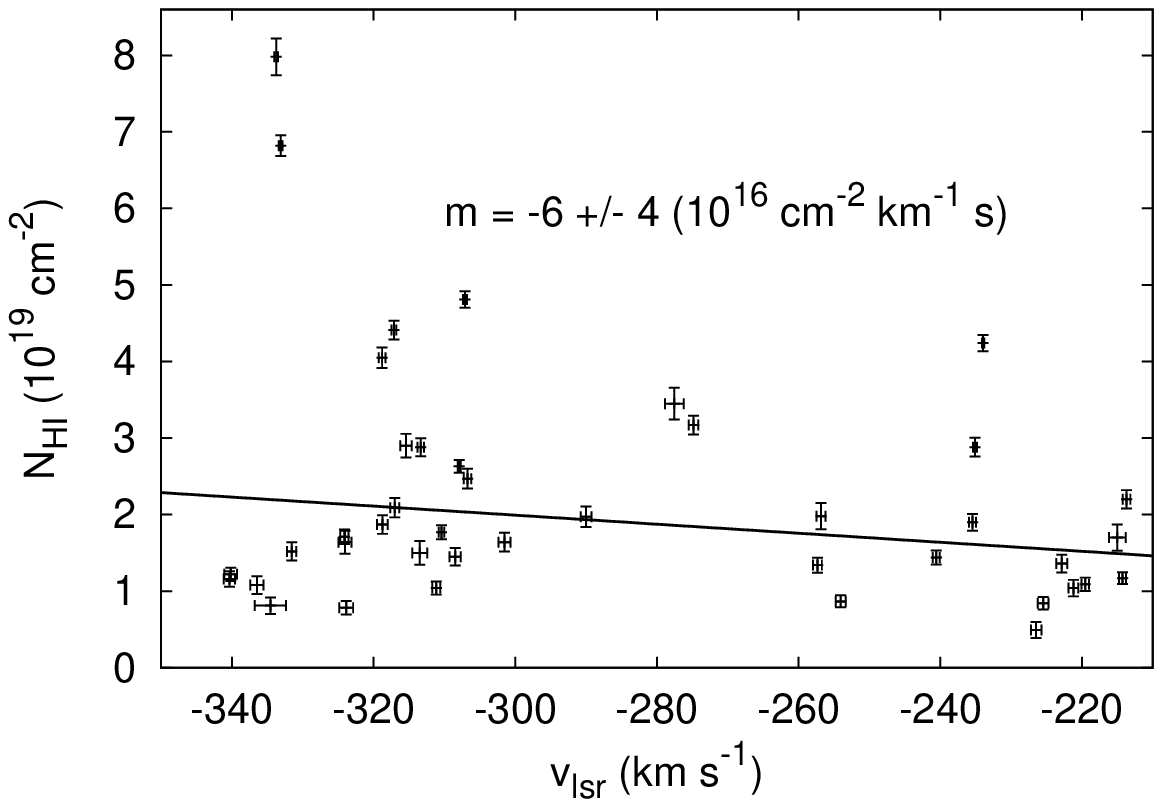}
\caption{Left panel: Histogram of the observed column densities. Right panel: Column density values vs. $v_\mathrm{lsr}$. We observe a small gradient, though the fit error is too high to make the result significant.}
\label{fighvcnhistats}
\end{figure}
In contrast to these earlier observations complex GCN turns out to be an association of tiny cloudlets mainly not even resolved  with the Effelsberg beam of 9\arcmin. In Fig.\,\ref{fighvcassociationplanes} we show two velocity planes out of the data cube revealing several compact clumps at LSR velocities in the range of $-340\ldots-210\,\mathrm{km\,s}^{-1}$. Even after averaging of several velocity planes no extended diffuse component becomes visible; see Fig.\,\ref{fighvcnhi}. (The EBHIS detection limit for a warm neutral component of $\Delta v=20\,\mathrm{km\,s}^{-1}$ is about $5\cdot10^{18}\,\mathrm{cm}^{-2}$.) The individual components have column densities in the range of $0.5\ldots8\cdot10^{19}\,\mathrm{cm}^{-2}$ (see Fig.\,\ref{fighvcnhistats} left panel), typical peak brightness temperatures of $0.5\,\mathrm{K}$, and have line widths of $15\ldots30\,\mathrm{km\,s}^{-1}$. The lack of a cold gas component is in agreement with previous measurements \citep[][using LAB data]{kalberla2006}, however, often the cold gas becomes visible as soon as the angular resolution of the observation is high enough \citep[e.g.,][]{wakker1997} which should be the case for EBHIS. We observe a small column density dependency on $v_\mathrm{lsr}$ of $-6\pm4\cdot10^{16}\,\mathrm{cm}^{-2}\,\mathrm{km}^{-1}\,\mathrm{s}$ though the error of the fit is high such that the dependency is not significant; see  Fig.\,\ref{fighvcnhistats} right panel.

\begin{figure}[!t]
\centering
\includegraphics[width=0.8\textwidth,bb=60 45 405 170,clip=]{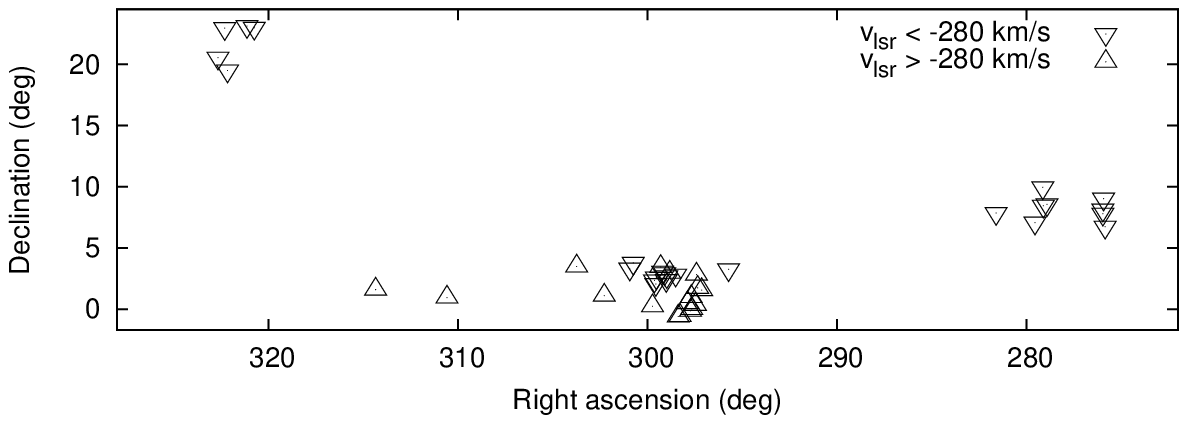}
\caption{Positions of the parametrized clumps. We identify four distinct clusters, three of which have velocities between $-340\ldots-300\,\mathrm{km\,s}^{-1}$ while the fourth cluster is at lower absolute velocities of about $-200\,\mathrm{km\,s}^{-1}$.}
\label{fighvccoords}
\end{figure}

Figure \ref{fighvcnhi} also shows that the clumps sometimes appear to form filaments or show head-tail-like structures giving hints on interactive processes like ram-pressure stripping or shocks. Note, that the cores in the filaments are generally still quite separated in velocity by more than $10\,\mathrm{km\,s}^{-1}$. The fraction of complex GCN located in the northern hemisphere can be separated into four clusters three of which lie at higher absolute velocities between $-340\ldots-300\,\mathrm{km\,s}^{-1}$ and the fourth cluster is at lower absolute velocities of about $-200\,\mathrm{km\,s}^{-1}$; see Fig.\,\ref{fighvccoords}. 

\begin{figure}[!t]
\centering
\includegraphics[width=0.8\textwidth,bb=60 54 405 195,clip=]{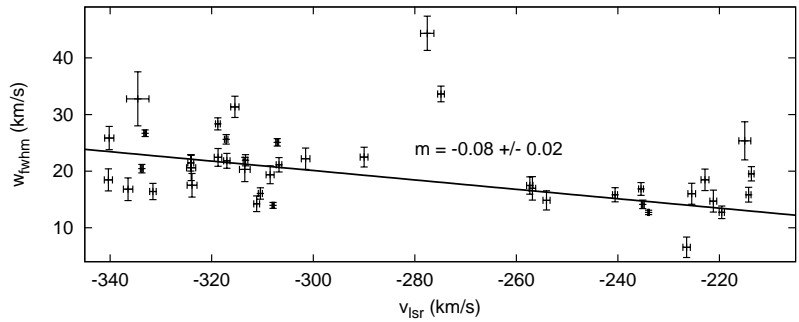}
\caption{Line widths of the clumps as a function of $v_\mathrm{lsr}$. There is a notable gradient of about $-1$.}
\label{fighvcwidths}
\end{figure}

While all detected clumps show no significant dependence of the column densities or peak temperatures on $v_\mathrm{lsr}$ there is a notable gradient of about $-1$ of the line widths with $v_\mathrm{lsr}$; see Fig.\,\ref{fighvcwidths}. Although this could simply be due to the fact, that the different line widths belong to the different clusters, the three clusters at higher negative velocities still have very similar line widths. If the observed trend was real it could be explained with the hypothesis that complex GCN is infalling gas accreting onto the Milky Way \citep[as previously suggested; see][and references therein]{wakker1991}. As the material has relative velocities about three times higher than the sound speed of the warm-hot ionized gas in the halo the high line widths and the lack of a diffuse extended component might indicate the presence of shocks compressing the infalling material, which would also explain the observed head-tail and filamentary structures. While the gas at higher negative velocities could not have had enough time to cool, the component at $-200\,\mathrm{km\,s}^{-1}$ might represent a later stage already decelerated and, hence, cooling processes had more time to produce the observed smaller line widths. Typical cooling times in the post-shock region of the observed clumps may be of the order of $10^6\ldots10^7\,\mathrm{yrs}$  similar to the free fall time of high-velocity clouds in the Milky Way halo as obtained from simulations \citep{quilis2001}.

\section{Summary}\label{secsummary}
We presented first results of the Milky Way data as observed with the EBHIS. The high angular resolution allowed us to resolve the HVC complex GCN into many small clumps. In contrast to measurements of many other HVC complexes/clouds where a cold compact gas component is embedded within a diffuse envelopes, neither we detect a warm extended component nor narrow velocity profiles. Together with the dependency of the line widths of the clumps on $v_\mathrm{lsr}$ the observations support the hypothesis that complex GCN is due to accretion of infalling material to the Milky Way.  We also plan to use GASS data to extend our analysis to the part of complex GCN located in the southern hemisphere. This will provide a further test to the mentioned hypothesis.

\acknowledgements We thank the Deutsche Forschungsgemeinschaft (DFG) for financial support under the research grant KE757/7-1. The shown results are based on observations with the 100-m telescope of the MPIfR (Max-Planck-Institut f\"ur Radioastronomie) at Effelsberg. Our work would not have been possible without the continuous support from the MPIfR and Effelsberg staff.

\bibliography{winkel_benjamin}

\end{document}